\documentclass[fleqn,10pt]{wlscirep}

\newcommand{\sizeA}{1.0\linewidth}
\graphicspath{{fig/}{./fig/}{.}}

\usepackage{epstopdf}

\usepackage{bm}
\usepackage{amsmath}
\usepackage{amssymb}
\usepackage[T1]{fontenc}

\title{Phase separations induced by a trapping potential in one-dimensional fermionic systems as a source of core-shell structures}

\author[1,2,*]{Agnieszka Cichy}
\author[3,$\ddagger$]{Konrad Jerzy Kapcia}
\author[3,$\dagger$]{Andrzej Ptok}
\affil[1]{Faculty of Physics, Adam Mickiewicz University, ul. Umultowska 85, PL-61-614 Pozna\'n, Poland}
\affil[2]{Institut f\"{u}r Physik, Johannes Gutenberg-Universit\"{a}t Mainz, Staudingerweg 9, D-55099 Mainz, Germany}
\affil[3]{Institute of Nuclear Physics, Polish Academy of Sciences, ul. W. E. Radzikowskiego 152, PL-31342 Krak\'{o}w, Poland}

\affil[ *]{agnieszkakujawa2311@gmail.com}

\affil[ $\ddagger$]{konrad.kapcia@ifj.edu.pl}

\affil[ $\dagger$]{aptok@mmj.pl}

\begin{abstract}
Ultracold fermionic gases in optical lattices give a great opportunity for creating different types of novel states.
One of them is phase separation induced by a trapping potential between different types of superfluid phases. 
The core-shell structures, occurring in systems with a trapping potential, are a good example of such separations.
The types and the sequences of phases which emerge in such structures can depend on spin-imbalance, shape of the trap and on-site interaction strength.
In this work, we investigate the properties of such structures within an attractive Fermi gas loaded in the optical lattice, in the presence of the trapping potential and their relations to the phase diagram of the homogeneous system.
Moreover, we show how external and internal parameters of the system and parameters of the trap influence their properties.
In particular, we show a possible occurrence of the core-shell structure in a system with a harmonic trap, containing the BCS and FFLO states.
Additionally, we find a spatial separation of two superfuild states in the system, one in the BCS limit as well as the other one in the tightly bound local pairs (BEC) regime.
\end{abstract}

\begin{document}

\flushbottom
\maketitle

\thispagestyle{empty}

\section*{Introduction}

Recently, an extensive progress in experimental techniques in ultracold quantum gases in optical lattices occurs.
These systems can be give as the greatest examples of practical realization of ``quantum simulators''~\cite{bloch.dalibard.08,giorgini.pitaevskii.08,guan.batchelor.13,georgescu.ashhab.14,dutta.gajda.15}.
Because of the fact that they are systems with fully controllable parameters, they open a new avenue to study fundamental phenomena of condensed matter physics. 
Moreover, such simulators can provide information about properties of physical systems in a context of different effects and mechanisms, which are difficult to observe in solid state materials due to their complexity.
An experimental realization of such systems with bosonic and fermionic gases in optical lattices have already been conducted~\cite{greiner.mandel.02,kohl.moritz.05,stoferle.moritz.06,jardens.strohmaier.08,jordens.tarruell.10}. 
It creates a very attractive field for further development.
Additionally, in these systems trapping potential and lattice geometry can be modified to study new exotic phases.
Hence, ultracold gases in optical lattices allow to simulate the well-known Hubbard model in various regimes of parameters~\cite{bloch.10}.

The lattice geometry is a fundamental characteristic of many-body systems and has large influence on their physical properties~\cite{ptok.17}.
Ultra-cold atomic gases give the opportunity of realization of systems with different geometries.
Experimentally, the geometry of the lattice can be changed by different spatial arrangement of laser beams~\cite{bloch.dalibard.08}.
Recently, the occurrence of antiferromagnetic spin correlations in the repulsive fermionic gas, in different lattice geometries of varying dimensionality, also including crossover configurations between different geometries, was investigated~\cite{corcovilos.baur.10,simon.bakr.11,hart.duarte.15,greif.jotzu.15,cheuk.nichols.16a,cheuk.nichols.16b,azurenko.chiu.17}.

Equally important is the better comprehension of unconventional phases that can appear in fermionic superfluids with population imbalance, the latter being the effect of a magnetic (Zeeman) field or the result of preparing a mixture with the desired composition.
In the weak coupling limit, states with nontrivial Cooper pairs can exist at large population imbalance~\cite{machida.mizushima.06,koponen.paananen.2008,liu.hu.08,cai.wang.11,ptok.cichy.17}. 
One of examples of such pairing is the Fulde--Ferrell--Larkin--Ovchinnikov (FFLO) state~\cite{FF,LO}, in which the Cooper pairs have non-zero total momentum as a result of pairing across the spin-split Fermi surface. 
The properties of this state attracted a lot of theoretical and experimental attention~\cite{casalbuoni.nardulli.04,matsuda.shimahara.10,beyer.wosnitza.13,ptok.kapcia.17,kinnunen.baarsma.18}.
Another example of unconventional coherent state is the homogeneous spin-polarized superconductivity with a gapless spectrum for the majority spin species~\cite{radzihovsky.sheehy.10}.
For this phase, the coexistence of  the normal and the superfluid states in the isotropic phase is characteristic.
This phase was firstly proposed by Sarma~\cite{sarma}  who studied the case of a superconductor in an external magnetic field within the BCS theory, neglecting orbital effects. 
He showed that self-consistent mean field solutions with gapless spectrum [$\Delta (h)$] are energetically unstable at $T=0$, in contrary to the fully gapped BCS solutions. 
On the other hand, a spin-polarized superconducting state can be stabilized by non-zero temperature.

Experimental studies of the spin--imbalanced fermionic gases give new possibilities for research in the field of condensed matter systems with strong correlations.
There are experimental evidences for the occurrence of 
{\it core-shell structure} --- an unpolarized superfluid core in the center of the trap surrounded by a polarized normal state~\cite{zwierlein.aboshaeer.05,zwierlein.schirotzek.06,partridge.li.06}.
The structure has been observed in the density profiles of trapped spin--imbalanced fermionic mixtures.
In this system a phase separation between these two states appears~\cite{liu.wilczek.03,sheehy.radzihovsky.06,diederix.gubbels.09,shin.zwierlein.06,partridge.li.06prl,lobo.recati.06,pilati.giorgini.08,giorgini.pitaevskii.08,bertaina.giorgini.09,valtolina.scazza.17}. 
Additionally, in the case of two-component fermionic gases in one-dimensional optical lattice, the exact thermodynamic Bethe ansatz solution indicates the occurrence of a mixed phase with two-shell structure~\cite{guan.batchelor.13}.
In such a state, a partially polarized superfluid core (the FFLO phase) is surrounded by a fully paired (BCS-like) or fully polarized (normal) phases~\cite{guan.batchelor.07,orso.07}.
Similar observation has been also performed in the case of a one-component trapped gas~\cite{hu.xiaji.07}. 
It is impotant to emphasize that the one-dimensional system is a good candidate for observing the FFLO phase because of a nesting effect, which makes the state much more robust than in the three-dimensional case~\cite{bloch.10,ptok.17}.

In this paper we show that the occurrence of core-shell structures is a consequence of the presence of inhomogeneity in the system, in particular, of the changes of chemical potential or magnetic field (or equivalently effective spin-imbalance) depending on the trapping potential and the short-range interactions between atoms.
Moreover, we provide an analysis according to which, depending on the attractive interaction, multiple core-shell structures can appear in the system, including different phases, such as the spatially homogeneous spin-unpolarized superconducting state, i.e., the BCS state, as well as the spatially inhomogeneous superconducting FFLO state.
We show that it is possible to prepare the system in such a way that one can observe the two different phases  separately in space and we study the influence of the trapping potential on such spatially separated phases.

\section*{Model and method}

In this paper, we study a one-dimensional system with the $s$-wave superconductivity.
The attractive Hubbard Hamiltonian (i.e., with on-site pairing, $U<0$) in presence of the magnetic field ($h$) has a following form:
\begin{eqnarray}
\label{eq.ham_real}
\mathcal{H} = \sum_{ \langle i,j \rangle \sigma } \left[ - t - ( \mu_{i} + \sigma h ) \delta_{ij} \right] c_{i\sigma}^{\dagger} c_{j\sigma} + U \sum_{i} n_{i\uparrow} n_{i\downarrow},
\end{eqnarray}
where $c_{i\sigma}^{\dagger}$ ($c_{i\sigma}$) denotes the creation (annihilation) operator of the particles at site $i$ and spin $\sigma= \{ \uparrow ,\downarrow \}$.
$t$ is the hopping integral between nearest-neighbor sites, $U < 0$ is the on-site pairing interaction, $h$ is the external  magnetic Zeeman field, while $\mu_{i}$ is an effective on-site chemical potential.
The interaction term is treated within the mean-field broken-symmetry approximation:
\begin{eqnarray}
n_{i\uparrow} n_{i\downarrow} = \Delta_{i}^{\ast} c_{i\downarrow} c_{i\uparrow} + \Delta_{i} c_{i\uparrow}^{\dagger} c_{i\downarrow}^{\dagger} - | \Delta_{i} |^{2},
\end{eqnarray}
where $\Delta_{i} = \langle c_{i\downarrow} c_{i\uparrow} \rangle$ is the {\it s}-wave superconducting order parameter (SOP).
Hence, the mean-field Hamiltonian in real space has the form:
\begin{eqnarray}
\mathcal{H}^{MF} = \sum_{ \langle i,j \rangle \sigma } \left[ - t - ( \mu_{i} + \sigma h ) \delta_{ij} \right] c_{i\sigma}^{\dagger} c_{j\sigma} + U \sum_{i} \left( \Delta_{i}^{\ast} c_{i\downarrow} c_{i\uparrow} + H.c. \right) - U \sum_{i} | \Delta_{i} |^{2}.
\end{eqnarray}

For a general case (i.e., for any distribution of $\mu_{i}$),
Hamiltonian~(\ref{eq.ham_real}) can be exactly diagonalized within the Bogoliubov--Valatin transformation:
\begin{eqnarray}
\label{eq.bvtransform} c_{i\sigma} = \sum_{n} \left( u_{in\sigma} \gamma_{n\sigma} - \sigma v_{in\sigma}^{\ast} \gamma_{n\bar{\sigma}}^{\dagger} \right) ,
\end{eqnarray}
where $\gamma_{n\sigma}$ and $\gamma_{n\sigma}^{\dagger}$ are the new {\it quasi}-particle fermionic operators, whereas ${\bm u}$ and ${\bm v}$ are the Bogoliubov--de~Gennes (BdG) eigenvectors.
One can obtain the self-consistent BdG equations in real space:
\begin{eqnarray}
\label{eq.bdgeq}
\sum_{j}
%%%%
\left(
\begin{array}{cc}
H_{ij\sigma} & \Delta_{ij} \\
\Delta_{ij}^{\ast} & -H_{ij\bar{\sigma}}^{\ast}
\end{array}
\right)
%%%%
\left( \begin{array}{c}
u_{jn\sigma} \\
v_{jn\bar{\sigma}}
\end{array} \right)
%%%%
=
%%%%
\mathcal{E}_{n\sigma}
%%%%
\left( \begin{array}{c}
u_{in\sigma} \\
v_{in\bar{\sigma}}
\end{array} \right)
%%%%
\end{eqnarray}
where $H_{ij\sigma} = - t \delta_{\langle i,j \rangle} - ( \mu_{i} + \sigma h ) \delta_{ij}$ is the single-particle Hamiltonian and $\Delta_{ij} = U \Delta_{i} \delta_{ij}$ are the on-site SOPs.
Using transformation~(\ref{eq.bvtransform}), the SOPs can be found as:
\begin{eqnarray}
\label{eq.sop_mom}
\Delta_{i} = \langle c_{i\downarrow} c_{i\uparrow} \rangle = \sum_{n} \left[ u_{in\uparrow} v_{in\downarrow}^{\ast} f( \mathcal{E}_{n\uparrow} ) - u_{in\downarrow} v_{in\uparrow}^{\ast} f ( - \mathcal{E}_{n\downarrow} ) \right] .
\end{eqnarray}
Equations~(\ref{eq.bdgeq}) can be solved self-consistently with respect to the distribution of $\Delta_{i}$. In this case, one can find the grand canonical potential for a given state as:
\begin{eqnarray}
\Omega \equiv - k_{B} T \sum_{n\sigma} \ln \left[ 1 + \exp \left( \frac{ - \mathcal{E}_{n\uparrow} }{ k_{B} T } \right) \right] - \sum_{i} \left( \mu_{i} + h + U | \Delta_{i} |^{2} \right) .
\end{eqnarray}
From several solutions of the BdG equations, only those with a minimal value of grand canonical potential $\Omega$ (at fixed $\mu$ and $h$) indicate a thermodynamically stable state of the system. 
In the absence of a trap (i.e., $\mu_{i} = \mu$ at each site), the distribution of $\Delta_{i}$ can be rewritten in momentum space, using the Fourier transform:
\begin{eqnarray}
\Delta_{i} = \frac{1}{N_{x}} \sum_{\bm q} \Delta_{\bm q} \exp \left( i {\bm R}_{i} \cdot {\bm q} \right) ,
\end{eqnarray}
where ${\bm q}$, which are restricted to the first Brillouin zone, are total momenta of the Cooper pairs.

\begin{figure}[!b]
\centering
\includegraphics[width=\sizeA]{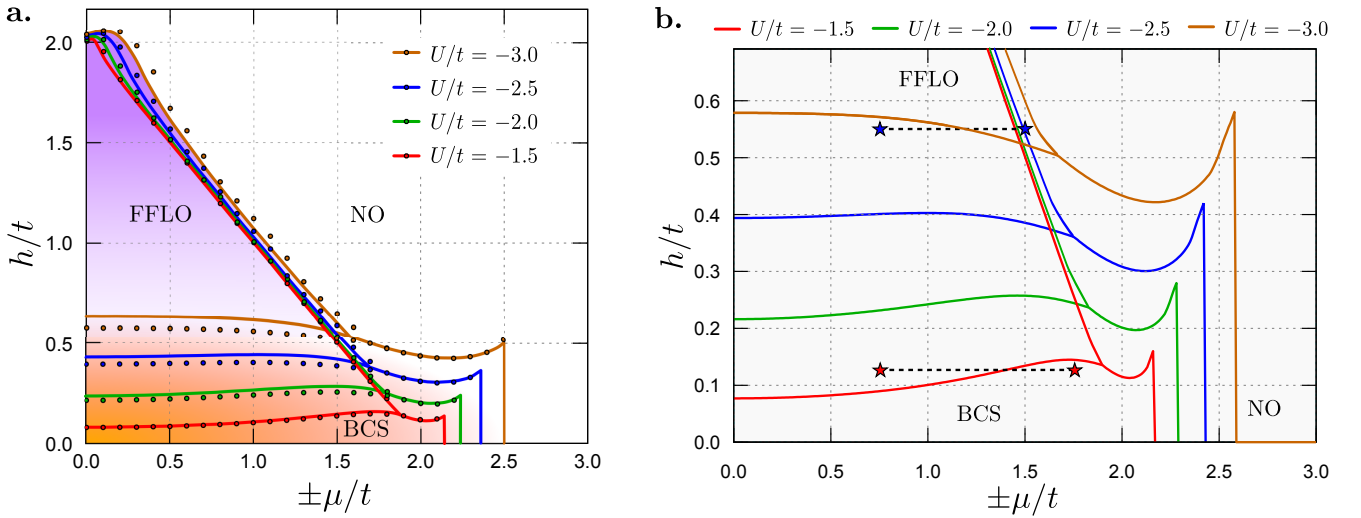}
\caption{
(a) The full $\mu$--$h$ ground state phase diagrams for the infinite homogeneous chain obtained for different values of on-site attraction $U/t$ (as labeled).
Labels indicate the regions of an occurrence of the following phases: NO -- normal phase; BCS -- unpolarized superconducting phase with ${\bm Q} = 0$; FFLO -- polarized superconducting phase with ${\bm Q} \neq 0$.
Solid lines and dots denote boundaries between the phases found from calculations in which only the simplified FF solutions are considered (cf. Ref.~\citeonline{ptok.cichy.17} and~\citeonline{ptok.cichy.17b}) as well as the full FFLO solutions are taken into account, respectively.
(b) A part of the full phase diagram in the vicinity of the BCS--FFLO transitions including only the full FFLO solutions.
The stars connected by the dashed lines denote sets $\mathcal{A}$ and $\mathcal{B}$ of the model parameters used in further calculations, included in the present work (cf. Figs.~\ref{fig.ps1} and~\ref{fig.ps2}).
\label{fig.df}
}
\end{figure}

\section*{Numerical results}

In this section, the obtained numerical results are presented.
First, the homogeneous system (without the trap) is investigated.
The  magnetic field $h$ versus chemical potential $\mu$ phase diagram is determined.
Next, we show that phase separation occurs in the presence of a harmonic trap.  
Such a phase separation induced by a trapping potential we will call \emph{(artificially) enforced} phase separation further in the text.
For chosen parameters (i.e., $\mu$ and trapping potential), the phase separation between different phases (e.g., a state with the BCS core and the FFLO shell as well as the FFLO core and the BCS shell) can be realized in the system.

\subsection*{Phase diagram: homogeneous system}

Numerical calculations presented in this section have been performed for a one-dimensional chain with the periodic boundary conditions and $N = 200$ sites.
For the homogeneous system, i.e., $\mu_{i} = \mu = \text{const.}$,  the  magnetic field $h$ versus chemical potential $\mu$ phase diagram is shown in Fig.~\ref{fig.df}.
The results for different values of $U/t$ are presented (cf. also Ref.~\citeonline{ptok.cichy.17} and~\citeonline{ptok.cichy.17b}).
In each case, they consist of three regions.
For low values of magnetic field, the conventional BCS phase is stable (with $\Delta_i=\text{const}$ and $\Delta_i\neq0$).
With increasing  $h$, one finds a discontinuous phase transition from the BCS to the FFLO phase (with $\Delta_i$ changing from site to site).
In the third region the normal (non-ordered, NO) phase is stable (with $\Delta_i=0$).
The transition from the BCS to the NO phase is continuous (for changing $h$ and fixed $\mu$) or discontinuous (at the vertical boundary in the phase diagram).
Notice also that for small $\mu/t$ and large $h/t$, the so-called $\eta$ phase can exist~\cite{yang.89} (the FFLO phase with maximal $q=\pi$, at least for the FF ansatz~\cite{ptok.cichy.17,ptok.cichy.17b}).
With increasing $U/t$, the region of the BCS phase extends, whereas the continuous FFLO--NO boundary weakly depends on $U/t$.
In Fig.~\ref{fig.df}(a), the solid lines correspond to the case in which only the FF state is considered, i.e., all Cooper pairs have the same momentum ${\bm q}$~\cite{FF}.
Including the fact that generally in the ``full'' FFLO state, Cooper pairs with different momenta can contribute (i.e., given by Eq.~(\ref{eq.sop_mom}), one gets that the region of the FFLO phase occurrence is slightly extended in comparison to the one obtained only for the FF ansatz.
This is an expected and well known relation between superconducting phases with different numbers of allowed ${\bm q}$, i.e., a phase with larger number of ${\bm q}$'s is more stable than the FF phase with only one ${\bm q}$~\cite{casalbuoni.nardulli.04,jiang.ye.07}.

It should be also mentioned that for low filling (i.e., for $| \mu | \sim 2 t$), the BCS--BEC crossover can be realized. 
The detailed discussion of this issue has been presented in Ref.~\citeonline{ptok.cichy.17}. 
In the context of the present work (particularly for the problem of the phase separation region in the system), the following property of the phase diagram is also very important.
Namely, depending on values of the magnetic field $h$ and on-site attraction $U$, with increasing $|\mu|$ the transitions from the BCS phase to the FFLO phase or from the FFLO phase to the BCS phase can occur [red and blue stars, respectively, in Fig.~\ref{fig.df}(b)].

However, it is important to keep in mind that the use of the mean-field approximation is generally restricted to the weak coupling limit and ground state properties.
The limitations of the mean-field method affect the one-dimensional system the most because the pair fluctuations become very important in this case~\cite{liu.hu.07}. 
For such system geometry, the nature of phase transition between the BCS and FFLO phases can be faultily predicted. 
However, the mean-field approximation can give some useful description in the weak and intermediate couplings, which are comparable with the Bethe ansatz results~\cite{liu.hu.07}.
While the mean-field FF-type calculations do not predict the correct type of the phase transition from the BCS phase to the FFLO phase, the self-consistent Bogoliubov--de~Gennes results are in a good agreement with those obtained from the Bethe ansatz. 
In the continuum model, the first order phase transition is simply an artifact of the FF-type calculation~\cite{liu.hu.07}.
This suggests a similar problem in the case of a lattice in the thermodynamic limit, when average particle concentration and lattice constant go to zero (i.e., $n\rightarrow0$ and $a\rightarrow0$ --- the dilute limit).

\subsection*{Phase separation and its types}

Generally, the phase separation is a state of the system where two or more uniform phases (e.g., those which have been defined previously) occur in different parts (so-called domains) of the system. 
In this section, we distinguish two different types of phase separation, which can emerge in the system, i.e.:
(a) a spontaneous phase separation, which can occur in a homogeneous system and 
(b) an artificially enforced phase separation, which can emerge in inhomogeneous system.
In the present work these cases correspond to the system with $\mu_i=\text{const}$ at each site and to the system with inhomogeneous spatial distribution of $\mu_i$, respectively.  
Below, we characterize briefly these two types of phase separations.

\subsubsection*{Spontaneous (macroscopic) phase separation}

The discontinuous transitions between two (homogeneous) phases in the diagram, as a function of $\mu$ are usually related to the discontinuous change of particle concentration from value $n_+$ to value $n_-$ ($n_+>n_-$).
Such a transition can be associated with the occurrence of the (macroscopic) phase separation in a defined range $n_-<n<n_+$ of particle concentration~\cite{arrigoni.strinati.91}.
In such a phase-separated state two domains with different particle concentrations $n_-$ and $n_+$ coexist (there can be also regions differing in the magnitude of the order parameter as well as thermodynamic phases).
In this approach, because of neglecting the interface energy at the boundaries of the domains, such states can exist only in the thermodynamic limit (i.e., when $N\rightarrow\infty$) \cite{kujawa.micnas.11,kapcia.robaszkiewicz.12,cichy.micnas.14,kapcia.czart.16,kapcia.baranski.17}. 
In a finite system, the interface energy can lead to an occurrence of states with other textures~\cite{coleman.yukalova.95,lorenzana.castellani.01.a,lorenzana.castellani.01.b,yukalov.yukalova.04,yukalov.yukalova.14}, besides the homogeneous states and the phase separated states discussed above.
Due to the fact that the BCS--FFLO, BCS--NO, and FFLO--NO boundaries in Fig.~\ref{fig.df}(a) can be discontinuous in the diagram as a function of $n$, one expects an occurrence of the following (macroscopic) phase separation regions: BCS/FFLO, BCS/NO, and FFLO/NO~\cite{ptok.cichy.17,ptok.cichy.17b}.
Notice that only the BCS--FFLO boundary is discontinuous for all model parameters.
To distinguish the phase separation from other states discussed above, we will call it a spontaneous one, because it can occur spontaneously in the homogeneous system.

\begin{figure*}[!t]
\centering
\includegraphics[width=\sizeA]{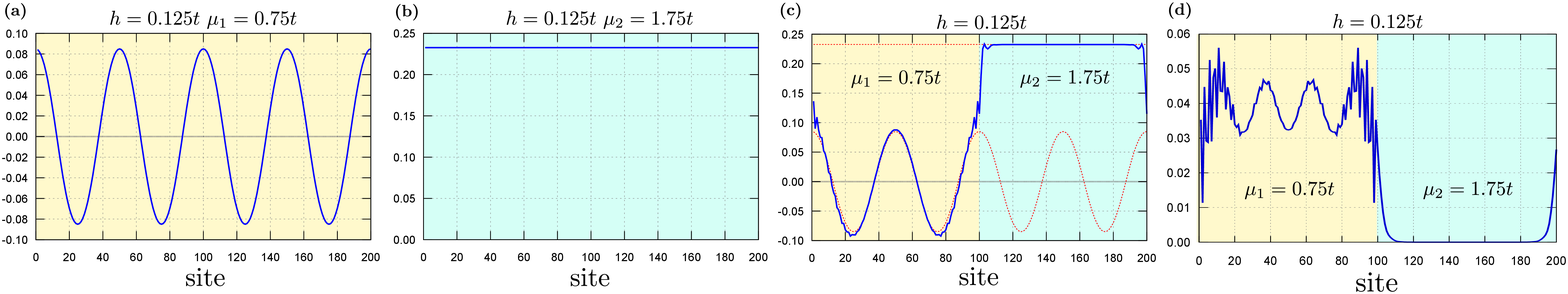}
\caption{
An example of the enforced phase separation of the FFLO and BCS phases.
Panels (a)-(c) show $U \Delta_{i}$ in the real space (i.e., as a function of lattice site $i$) for $U = -1.5 t$ and $h=0.125t$ (set $\mathcal{A}$ of the model parameters).
Panels (a) and (b) are obtained for the homogeneous system in which the FFLO phase ($\mu=\mu_1=0.75t$) and the BCS phase ($\mu=\mu_2=1.75t$) exist, respectively [i.e., for the model parameters denoted by the red stars in Fig.~\ref{fig.df}(b)].
(c) The inhomogeneous state (the enforced phase separation) for the system with two different chemical potentials: $\mu_i=\mu_1$ for the left half of the system, and $\mu_i=\mu_2$ for the right half.
The red doted lines present the results from panels (a) and (b), for the comparison.
(d) The local polarizations are shown in panel (c). The background colors (yellow and blue) correspond to the stable phases for
fixed $\mu$ (lower value $\mu_1$ and higher value $\mu_2$,
respectively).
\label{fig.ps1}
}
\end{figure*}

\begin{figure*}[!t]
\centering
\includegraphics[width=\sizeA]{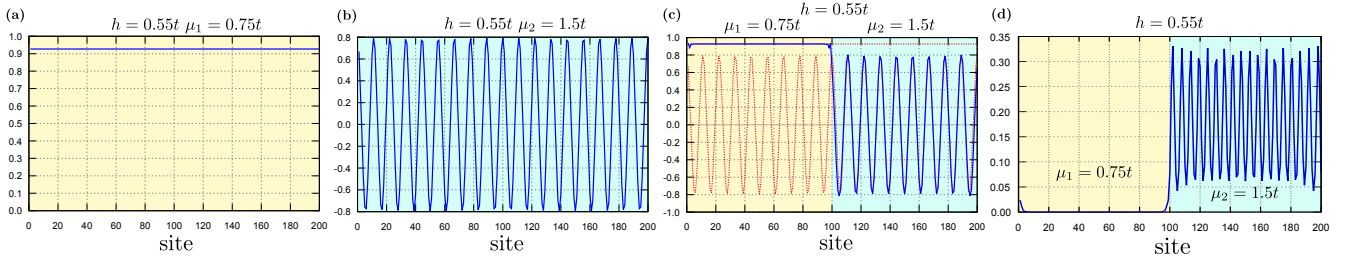}
\caption{
The same as in Fig.~\ref{fig.ps1} but for $U = -3.0 t$, $h = 0.55 t$ (set $\mathcal{B}$ of the parameters) and $\mu_1=0.75t$ and $\mu_1=1.5t$ [i.e., the model parameters denoted by the blue stars in Fig.~\ref{fig.df}(b)]. The background colors (yellow and blue) correspond to the stable phases for
fixed $\mu$ (lower value $\mu_1$ and higher value $\mu_2$,
respectively).
\label{fig.ps2}
}
\end{figure*}

\subsubsection*{Artificially enforced phase separation: a model example}

From the analysis of the phase diagram and shapes of boundaries between the phases, it is clear that for $U/t=-1.5$ and $h/t=0.125$ (set $\mathcal{A}$ of the model parameters), the phase with the lowest energy is (a) the FFLO phase at $\mu=\mu_1=0.75t$ and (b) the BCS phase at $\mu=\mu_2=1.75t$ [the red stars in Fig.~\ref{fig.df}(b)].
The SOP at each site for these two solutions is presented in Fig.~\ref{fig.ps1}(a) and Fig.~\ref{fig.ps1}(b).
Analogously, we also choose the following parameters for $U/t=-3.0$ and $h/t=0.55$ [set $\mathcal{B}$ of the parameters, the blue stars in Fig.~\ref{fig.df}(b)]: $\mu_1=0.75t$ (the BCS phase) and $\mu_2=1.5t$ (the FFLO phase), and present the local dependence of the SOP in Fig.~\ref{fig.ps2}(a) and Fig.~\ref{fig.ps2}(b).
One can notice that in the FFLO phase $\Delta_i$ changes from site to site periodically, whereas for the BCS solution $\Delta_i$ is homogeneous in the whole system.

Now, we investigate the system with a particular distribution of the chemical potential: $\mu_i=\mu_1$ in one half of system (favoring the FFLO phase or the BCS phase for set $\mathcal{A}$ or $\mathcal{B}$, respectively) and $\mu_i=\mu_2$ in the other half (favoring the BCS phase or the FFLO phase for set $\mathcal{A}$ or $\mathcal{B}$, respectively).
After solving the self-consistent set of the BdG equations in real space for such a system with $N=200$ sites (and with open boundary conditions), one gets that solutions obtained in parts of the system resemble the ones for the homogeneous system [Fig.~\ref{fig.ps1}(c) and Fig.~\ref{fig.ps2}(c)].
In such a case, we have a coexistence of two homogeneous solutions, which are spatially separated.
Notice that small changes (in comparison to the solutions for the homogeneous system in each part of the system) are distinguishable only in the neighborhood of interfaces between two such defined domains.
Thus, one can conclude that the effects associated to the interfaces are almost irrelevant for the state under consideration.
Moreover, the spin polarization (defined as the difference of concentrations of particles with spin up and spin down, $m_i=n_{i\uparrow}-n_{i\downarrow}$) as a function of site $i$ in these inhomogeneous states is shown in Fig.~\ref{fig.ps1}(d) and Fig.~\ref{fig.ps2}(d), for $\mathcal{A}$ and $\mathcal{B}$, respectively.
As can be expected, the spin polarization has a modulation, which is two times faster than that of the SOP.
Additionally, maximal values of the spin polarization are located at the nodal points of the SOP~\cite{yanase.sigrist.09,loh.trivedi.10}.
This is a consequence of the interplay between the unpaired polarized particles and those in the superconducting state (i.e., the Cooper pairs).
A small number of pairs (at sites with $\Delta_i\approx 0$) supports the occurrence of polarized states (i.e., a large value of $m_i$). 
Note also that there is no coexistence of the superconducting and magnetic ordering in the BCS phase.

The nature of this inhomogeneous state is, however, different than the origin of the spontaneous (macroscopic) phase separation discussed in the previous point.
In the present setup, the system parameters are inhomogeneous (different values of chemical potential in two parts of the system).
We will call such separated state an (artificially) enforced phase separation in contrary to the spontaneous phase separation, which could occur in the homogeneous system.

Note, that even if $\mu_1=\mu_2=\mu_c$ (where $\mu_c$ is a critical value of the chemical potential at the FFLO--BCS boundary for particular $U/t$ and $h/t$), due to the fact that the interface has small but still finite energy, the finite system cannot exhibit spontaneous phase separation.
In such a case, the whole system is in the FFLO phase or in the BCS phase (both states have equal energy and both solutions correspond to local minima of the grand canonical potential).

\subsection*{System with a trap}

After studying the model systems, let us investigate more realistic situations, in which the on-site potential $\mu_i$ changes from site to site.
An example of experimental realization of such systems are ultracold atomic gases in optical lattices.
In the following, we consider two types of traps: (i) a linear trap with $\mu_{i} = V_{0} | r_{i} - r_{c} |$ and (ii) a harmonic trap with $\mu_{i} = V_{0} ( r_{i} - r_{c} )^{2}$, where $r_{i}$ is a location of $i$-th site, whereas $r_{c}$ is a location of the center of the trap (i.e., $i=N/2=500$).
In this part, we performed the calculations for $N=1000$ sites with open boundary conditions.
$V_0$ was chosen in such a way to ensure a value much larger than the chemical potential corresponding to the BCS-NO transition in the homogenous system at $h=0$ and fixed $U$.

\begin{figure}[!t]
\centering
\includegraphics[width=\sizeA]{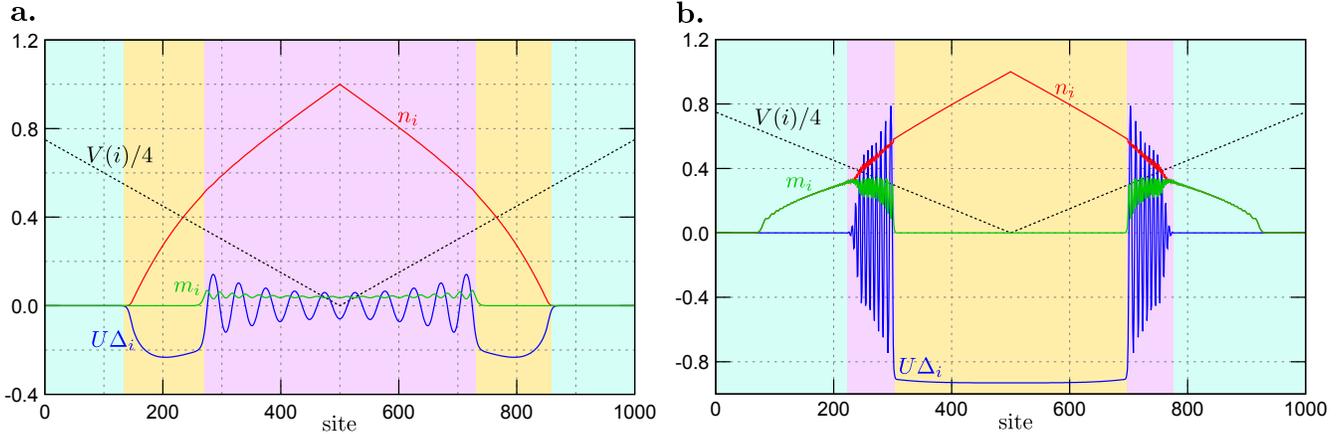}
\caption{
SOP $U \Delta_{i}$, magnetization $m_i$, and particle concentration $n_i$ (as labeled, solid lines) in real space (as a function of lattice site $i$) in the presence of a linear trap, for two different values of on-site attraction and external magnetic field: (a) $U/t=-1.5$ and $h=0.125t$ (set $\mathcal{A}$ of the parameters); and (b) $U/t=-3.0$ and $h=0.55t$ (set $\mathcal{B}$ of the parameters).
The dependence of (local) chemical potential $\mu_i$ is marked with the dotted line. The background colors (blue, yellow and pink) indicate
the boundaries between the phases (NO, BCS and FFLO, respectively).
\label{fig.linear}
}
\end{figure}

\begin{figure}[!b]
\centering
\includegraphics[width=\sizeA]{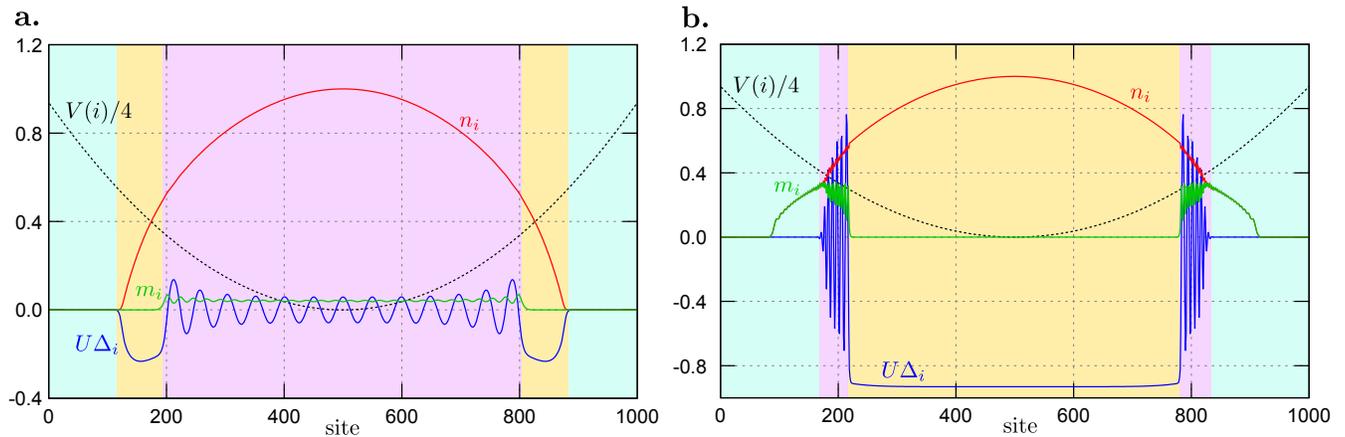}
\caption{
SOP $U \Delta_{i}$, magnetization $m_i$, and particle concentration $n_i$ (as labeled, solid lines) in the real space (as a function of lattice site $i$) in a presence of the harmonic trap, for two different values of on-site attraction and external magnetic field: (a) $U/t=-1.5$ and $h=0.125t$ (set $\mathcal{A}$ of the parameters); and (b) $U/t=-3.0$ and $h=0.55t$ (set $\mathcal{B}$ of the parameters).
The dependence of (local) chemical potential $\mu_i$ is marked with the dotted line. The background colors (blue, yellow and pink) indicate
the boundaries between the phases (NO, BCS and FFLO, respectively).
\label{fig.harmonic}
}
\end{figure}

The  self-consistent solutions of the BdG equations for the case of a linear trap are presented in Fig.~\ref{fig.linear}.
In this case, we show SOP $U \Delta_{i}$, magnetization $m_i$, particle concentration $n_i$, and the trapping potential $\mu_i \equiv V(i)  = V_{0} | r_{i} - r_{c} |$ (with $V_0 = 6.0 t / N$) as a function of the lattice site $i$ (blue, green, red and dashed black lines, respectively).
Due to the fact that $\mu_i$ changes from site to site, $n_i$ changes at each site and one does not observe the solutions with, e.g., constant value of $\Delta_i$, corresponding to the BCS phase (but the region without oscillations of the SOP can be identified as corresponding to this phase, see below).
However, one can indicate the regions along the chain which correspond to the phases indicated in Fig.~\ref{fig.df}(a).
Namely, starting from the left side of Fig.~\ref{fig.linear}(a) (for $U/t=-1.5$ and $h=0.125t$, set $\mathcal{A}$), one can find a region of the NO (empty) phase, where $\Delta_i=0$ (and $n_i=0$).
Next, there is the BCS region, where $\Delta_i\neq0$ exhibits no oscillations (and $m_i=0$).
The oscillations of $\Delta_i$ and nonzero magnetization $m_i\neq0$ are clearly visible in the center of the system.
Such a behaviour of $\Delta_i$ indicates the occurrence of the FFLO phase in this part of the system.
Hence, in the linear trapping potential, the FFLO phase in the core (i.e., in the center of the system) is surrounded by the shell of the BCS states.
For $U/t=-3.0$ and $h=0.55t$ (set $\mathcal{B}$), the situation is analogous, but going to the center of the trap, the following sequence of regions is presented: the NO phase [with $n_i=0$ (vacuum state), as well as filled state with $n_i\neq0$], the FFLO phase, and the BCS phase (as a core).
The NO phase with $n_i\neq0$ is fully polarized (i.e., $m_i/n_{i}=1$).

Note, that the location $i$ of the boundaries between the regions along the chain corresponds, approximately, to the values of $\mu_i$ for which the phase transition occurs in the homogeneous system [cf. Fig.~\ref{fig.df}(b)].
It is attributed to the fact that the interactions in the system are short-ranged and the interface energy is relatively small.
Moreover, for set $\mathcal{A}$, only the NO phase with $n=0$ can exist, whereas for set $\mathcal{B}$ also the NO phase with $n\neq0$  is possible (cf. Ref.~\citeonline{ptok.cichy.17}).

The results for a harmonic trap, presented in Fig.~\ref{fig.harmonic}, do not differ qualitatively from those obtained for a linear trap.
We take $\mu_{i} \equiv V(i) = V_{0} ( r_{i} - r_{c} )^{2}$, with $V_{0} = 4.0 t / (N/2)^2$.
Such chosen values correspond to a relatively flat trap.
The sequences of the regions for both sets of the model parameters ($\mathcal{A}$ and $\mathcal{B}$) are NO--BCS--FFLO and NO--FFLO--BCS, respectively, with spatially extended regions in the center (cores) and spatially shrunken external regions (shells).

\begin{figure}[!t]
\centering
\includegraphics[width=\sizeA]{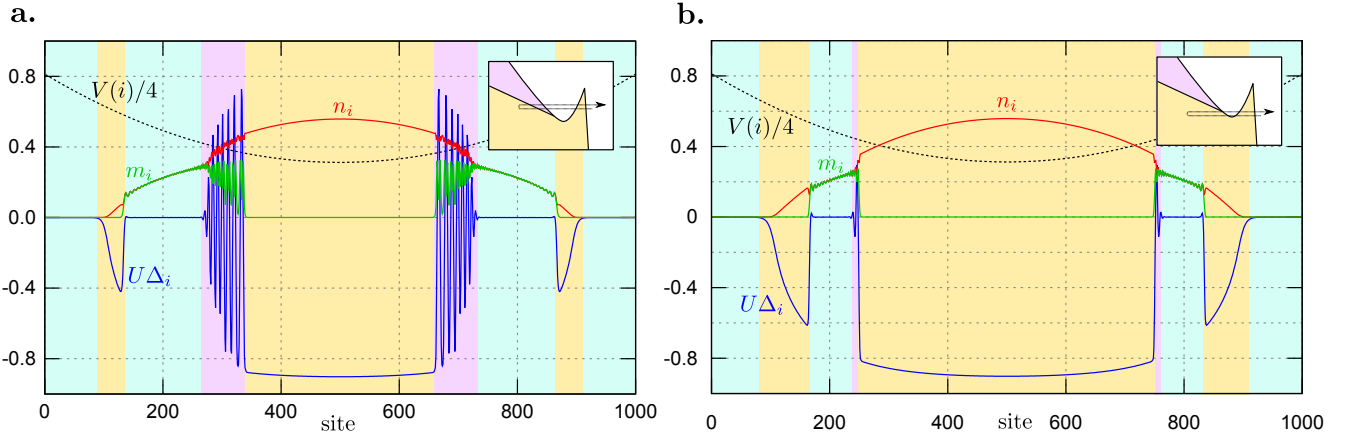}
\caption{
SOP $U \Delta_{i}$, magnetization $m_i$, and particle concentration $n_i$ (as labeled, solid lines) in the real space (as a function of lattice site $i$) in a presence of the shifted harmonic trap.
The results obtained  for (a) $U = -3.0 t$ and $h = 0.45 t$;  and (b)  $U = -3.0 t$ and $h = 0.375 t$.
The dependence of (local) chemical potential $\mu_i$ is marked with the dotted line.
Insets present schematically the region of the phase diagram for homogeneous systems corresponding to the range of $\mu_{i}$ changes [cf. with Fig.~\ref{fig.df}(b)]. The background colors (blue, yellow and pink) indicate
the boundaries between the phases (NO, BCS and FFLO, respectively).
\label{fig.harmonic_shif}
}
\end{figure}

The results presented above show that in the system with a trap at each site, the state of a homogeneous system corresponding to $\mu=\mu_i$ is approximately realized. 
The trap shapes considered above are chosen in such a way to change $\mu_i \equiv V(i)$ from a range that all of possible phases occurring in the system are present, at fixed $h$ cf. Fig.~\ref{fig.df} (from the NO phase at large $|\mu|$ at the boundaries of the system, i.e., at $i=0,N-1$, to the half-filled phases at $\mu=0$ in the center, i.e., at $i=N/2$). 
Thus, in other words, all phases for a homogeneous system at fixed $\mu$ can be realized in the setup with the trap. 
However, by modification of parameters of the trap for instance, one can realize a system which corresponds only to a part of the phase diagram (i.e., some limited part of $\mu$ range). 
A sample of the results is shown in Fig.~\ref{fig.harmonic_shif}.
They are obtained for the system with a harmonic trap in the form of $\mu_{i} = V_{0} ( r_{i} - r_{c} )^{2} + \bar{V}$, with $V_{0} = 2.0 t / (N/2)^2$ and $\bar{V} = 1.25 t$.
In this case, the minimum value of the trapping potential is shifted to some finite value, i.e., $\mu_{N/2} = \bar{V}$ (black doted line).
The chosen parameters allow to realize a part of the phase diagram in which two superconducting domains are separated by the NO state [cf. Fig.~\ref{fig.df}(b)].
Depending on a fixed value of magnetic field $h$, the states with BCS or BCS-FFLO core surrounded by the BCS shell, separated by the NO sub-shell-part can be realized in the system.

The results shown in Fig.~\ref{fig.harmonic_shif} correspond to a part of the phase diagram [cf. Fig.~\ref{fig.df}(b)], where the BCS--BEC crossover is realized. 
One should notice that the system under consideration is in a trap, which can change the state at the site with $\mu_i$ in comparison to the homogeneous solution with $\mu=\mu_i$. 
Indeed, numerical results show that this state can be realized outside the BCS core state [Fig.~\ref{fig.harmonic_shif}(a)], for $h$ smaller than the one at which the FFLO phase occurs in a homogeneous system.
The realization of the two BCS-like domains, separated by the NO state, is possible by the decrease of $h$ [Fig.~\ref{fig.harmonic_shif}(b)].
At the boundary of the internal BCS core, one observes the FFLO-type gap oscillation with changing of its sign~\cite{castorina.grasso.05,iskin.williams.08,bakhtiari.leskinen.08}.
It is important to emphasize that, as a consequence of  low filling at the external BCS shell state, we expect realization of the BEC condensate (the tightly bound local pairs region) of the Cooper pairs. 
According to the {\it Leggett criterion}~\cite{leggett.80}, the BEC (i.e., Bose--Einstein condensate) begins when the effective chemical potential is smaller than the lower band edge.

\section*{Summary}

In conclusion, we have shown that properties of the system in the presence of a trap are strongly associated to the phase diagram of the homogeneous system (in the absence of a trap). 
It is clearly seen from the spatial dependence of different quantities in the system with linear and harmonic trap (Figs.~\ref{fig.linear} and \ref{fig.harmonic}, respectively) and provides a confirmation about the validity of so-called local density approximation, that is very often employed in theoretical calculations for trapped system starting from previous results for homogeneous systems \cite{heidrichmeisner.orso.10,hu.xiaji.07,liu.hu.08}.
We have shown that in such a system the core-shell structures can be created due to the trapping potential.
For chosen parameters, the  order of states realized in such structures corresponds to the sequence of phases occurring in the phase diagram for the homogeneous system (Fig.~\ref{fig.df}).
The states occurring in a particular sequence depend mainly on values of model parameters (mainly $h$ and $U$). 
Generally, the shape of the trapping potential does not change the structure of phases occurrence.
Moreover, even in the same type of trap, for different values of the model parameters (i.e., interaction $U$ or magnetic field $h$), the structure with the BCS core and  the FFLO shell as well as with the FFLO core and the BCS shell can be realized (depending on $U$ and $h$).

An increase of the on-site interaction at low filling can lead to the BCS--BEC crossover in the system.
By tuning of the trap parameters, we can realize the part of the phase diagram in which this crossover occurs.
Thus, one can obtain the more complex core-shell structures.
In particular, one can obtain two BCS states separated by the NO state (e.g., Fig.~\ref{fig.harmonic_shif}).
Due to the low filling, at the outer BCS shell, the BEC should emerge in the system.

In this work, we have shown that different core-shell structures can occur in a trapped system.
These structures are examples of so-called (artificially) enforced phase separation, occurring in spatially inhomogeneous systems, whose origin is different than that of the macroscopic phase separation in homogeneous systems.
It is important to emphasize that such experimental setup with a trap allows to investigate phase diagrams of homogeneous systems with short-range interactions.
The theoretical prediction presented in this work should be realizable experimentally in a relatively simple way.

\bibliography{biblio}

\section*{Acknowledgements}

We thank Krzysztof Cichy for careful reading of the manuscript, valuable comments and discussions. 
A.P. is grateful to Laboratoire de Physique des Solides (CNRS, Universit\'{e} Paris-Sud) for hospitality during a part of the work.
This work was supported by the National Science Centre (Narodowe Centrum Nauki, NCN, Poland) under grants nos.: 
UMO-2017/24/C/ST3/00357 (A.C.), 
UMO-2017/24/C/ST3/00276 (K.J.K.), 
and 
UMO-2017/25/B/ST3/02586 (A.P.).

\section*{Author contributions statement}

A.P. initialized and coordinated the project.
A.P. performed numerical calculation.
All authors discussed the results.
A.C. and K.J.K. prepared the first version of the manuscript.
All authors contributed to the final version of the manuscript and all of them reviewed and accepted it.

\section*{Competing interests}

The authors declare no competing interests.

\end{document}